\documentclass[fleqn,usenatbib]{mnras}

\usepackage{newtxtext,newtxmath}

\usepackage[T1]{fontenc}

\DeclareRobustCommand{\VAN}[3]{#2}
\let\VANthebibliography\thebibliography
\def\thebibliography{\DeclareRobustCommand{\VAN}[3]{##3}\VANthebibliography}

\usepackage{graphicx}	
\usepackage{amsmath}	

\title[Simulated JWST galaxy damping wings]{JWST observations of galaxy damping wings during reionization interpreted with cosmological simulations}

\author[L. C. Keating et al.]{Laura C. Keating$^{1}$\thanks{laura.keating@ed.ac.uk},  James S. Bolton$^{2}$, Fergus Cullen$^{1}$, Martin G. Haehnelt$^{3}$, Ewald Puchwein$^{4}$ \newauthor
and Girish Kulkarni$^{5}$
\\
$^{1}$Institute for Astronomy, University of Edinburgh, Blackford Hill, Edinburgh, EH9 3HJ, UK\\
$^{2}$School of Physics and Astronomy, University of Nottingham, University Park, Nottingham, NG7 2RD, UK\\
$^{3}$Kavli Institute for Cosmology and Institute of Astronomy, Madingley Road, Cambridge, CB3 0HA, UK\\
$^{4}$Leibniz-Institut f\"ur Astrophysik Potsdam, An der Sternwarte 16, 14482 Potsdam, Germany\\
$^{5}$Tata Institute of Fundamental Research, Homi Bhabha Road, Mumbai 400005, India\\
}

\date{Accepted ---. Received ---; in original form ---}

\pubyear{2023}

\begin{document}
\label{firstpage}
\pagerange{\pageref{firstpage}--\pageref{lastpage}}
\maketitle

\begin{abstract}
Spectra of the highest redshift galaxies taken with JWST are now allowing us to see into the heart of the reionization epoch. Many of these observed galaxies exhibit strong damping wing absorption redward of their Lyman-$\alpha$ emission. These observations have been used to measure the redshift evolution of the neutral fraction of the intergalactic medium and sizes of ionized bubbles. However, these estimates have been made using a simple analytic model for the intergalactic damping wing. We explore the recent observations with models of inhomogeneous reionization from the Sherwood-Relics simulation suite. We carry out a comparison between the damping wings calculated from the simulations and from the analytic model. We find that although the agreement is good on the red side of the Lyman-$\alpha$ emission, there is a discrepancy on the blue side due to residual neutral hydrogen present in the simulations, which saturates the intergalactic absorption. For this reason, we find that it is difficult to reproduce the claimed observations of large bubble sizes at $z \sim 7$, which are driven by a detection of transmitted flux blueward of the Lyman-$\alpha$ emission. We suggest instead that the observations can be explained by a model with smaller ionized bubbles and larger intrinsic Lyman-$\alpha$ emission from the host galaxy.
\end{abstract}




\section{Introduction}

Absorption lines observed in the spectra of luminous objects, blueward of the Lyman-$\alpha$ (Ly$\alpha$) wavelength at the redshift of the source, indicate the presence of intergalactic neutral hydrogen. At redshifts approaching the epoch of reionization, this absorption is observed to saturate and regions with no significant transmission are detected on scales of tens of comoving Mpc \citep{becker2015,zhu2021}. However, it is difficult to use these observations to infer the presence of completely neutral gas, as for gas at the mean cosmic density, complete saturation will occur already at a volume-weighted average neutral hydrogen fraction $\langle x_{\rm HI} \rangle_{\rm v} \sim 10^{-5}$ \citep{mcquinn2016}. 

More compelling evidence for a significantly neutral intergalactic medium (IGM) is an observation of Ly$\alpha$ absorption that extends to wavelengths redward of the source redshift \citep{miraldaescude1998}. When the diffuse IGM has a high volume filling factor of neutral gas, its optical depth becomes large enough that the absorption profile is dominated by the natural line broadening described by the Lorentzian component of the Voigt profile  \citep{meiksin2009}. This means that there is a reasonable probability for photons to scatter in the wings of the line, even several thousands of kilometres away from line centre. This results in an absorption profile known as a damping wing. The strength of the damping wing is proportional to the optical depth of the diffuse IGM and hence the volume-averaged neutral fraction \citep{gunn1965,chen2023}, allowing for constraints on the reionization history from individual bright objects.

Multiple quasars observed at $z=7$ and above display the signature of IGM damping wings \citep{mortlock2011,banados2018,wang2020,yang2020}. By comparing the observed damping wings with theoretical models that also take local ionization by the quasar into account, the neutral fraction of the average IGM at the source redshift can be inferred \citep{bolton2011,greig2017,greig2019,davies2018}. A challenge in these measurements is the determination of the intrinsic Ly$\alpha$ emission of the quasar, with different reconstructions leading to different constraints on the neutral fraction \citep[e.g.,][]{greig2022}. Despite this, there is consensus from the different analyses of the observed damping wings that reionization is incomplete above $z=7$ \citep{fan2022}. The constraints are further limited by the redshifts of the most distant quasars currently known, although this is expected to improve with the Euclid wide survey, with more than 60 $z > 8$ quasars with magnitude $H<24$ predicted to be discovered \citep{schindler2023}.

The presence of damping wings in the spectra of long-duration gamma-ray burst (GRB) afterglows can also be used to constrain the timing of reionization \citep{totani2006,chornock2013,hartoog2015}. This has some advantages over studies with quasars \citep{barkana2004}. The spectra of GRB afterglows close to the Ly$\alpha$ wavelength can be described as a power law, sidestepping the issue of reconstructing an intrinsic emission line profile. The hosts of GRBs are further expected to be star-forming galaxies, which are more numerous than quasars and which likely live in less biased environments of the Universe.  The main disadvantage is that these GRB afterglows frequently show evidence for damped Ly$\alpha$ systems (DLAs) at the redshift of the host galaxy \citep[e.g.,][]{jensen2001}, which will also produce a damping wing that may eclipse the intergalactic signal. There are also relatively few high signal-to-noise GRB afterglow spectra currently available, but detections by future missions are predicted to provide competitive constraints on the evolution of the IGM neutral fraction \citep{lidz2021,tanvir2021}.

The impact of IGM damping wings can also be detected statistically from populations of galaxies. The luminosity function of Ly$\alpha$ emitting galaxies is observed to decline above $z=6$ more rapidly than expected from comparison with UV luminosity functions \citep{konno2014, konno2018}, indicating that absorption from the IGM may be obscuring the Ly$\alpha$ emission of the galaxies. This is complemented by the declining fraction of Lyman-break galaxies that show Ly$\alpha$ emission above $z=6$ \citep{stark2010, pentericci2011, pentericci2014, schenker2012, schenker2014}. Through comparison with theoretical models, these observations can be translated into constraints on the progress of reionization \citep{mason2018, mason2019, hoag2019, jung2020, bolan2022, wold2022, jones2023}. Detections of Ly$\alpha$ emission in individual galaxies can also be used to estimate the sizes of ionized bubbles surrounding the host galaxies \citep{mason2020,whitler2023,witstok2023}.

JWST is now opening up a new window in the study of IGM damping wings in galaxies, with damping wings being routinely observed in the spectra of individual galaxies \citep{curtislake2023}. By comparing the strength of these damping wings with the analytic model of \citet{miraldaescude1998}, the spectra can be used to place constraints on the IGM neutral fraction \citep{curtislake2023, hsiao2023}. \citet{umeda2023} used the same model to fit the strength of damping wings in stacks of galaxy spectra in different redshift bins, placing constraints on the evolution of the volume-weighted average neutral fraction and ionized bubble size out to $z=12$. They found evidence for an IGM neutral fraction that is increasing with redshift, and that is consistent with empirical models of galaxy-driven reionization \citep{ishigaki2018, finkelstein2019, naidu2020}. Interestingly, the bubble sizes that \citet{umeda2023} measured were towards the larger end of what is predicted by theoretical models \citep{lu2023}. However, as with the GRB afterglows, the interpretation of these results may be complicated by DLAs associated with the host galaxies. Indeed, \citet{heintz2023} recently found evidence for three $z > 8.8$ galaxies hosting DLAs with column densities $N_{\rm HI} > 10^{22} \, \rm{cm}^{-2}$.

Interpretation of these observations requires a model that captures all of the relevant physics, such as residual neutral gas in ionized regions of the IGM \citep{MesingerHaiman2004, laursen2011, bolton2013, mesinger2015, mason2020}, the impact of infalling gas onto the host halo \citep{dijkstra2007, sadoun2017, weinberger2018, park2021} and the inhomogeneity of reionization \citep{mesinger2008, mcquinn2008, garel2021, gronke2021, smith2022, qin2022, chen2023}. In this paper, we explore recent JWST observations of IGM damping wings in the context of these issues, analysing damping wings constructed from lines of sight through simulations of inhomogeneous reionization from the Sherwood-Relics simulation suite \citep{puchwein2023}. Specifically, we critically evaluate the use of the \citet{miraldaescude1998} model to constrain reionization with recent JWST observations.

We describe our models of the IGM damping wing in Section \ref{section:models}. In Section \ref{section:compare_model}, we compare the results of the damping wings constructed from the reionization simulations to the commonly used \citet{miraldaescude1998} model. Section \ref{section:compare_observations} presents a comparison of our results with damping wings observed in JWST galaxies, focusing specifically on the recent results of \citet{umeda2023} and \citet{heintz2023}. Finally, in Section \ref{section:conclusions}, we summarise our results and present our conclusions. Throughout the paper, we assume the \citet{planck2014} cosmological parameters, namely $\Omega_{\rm m} = 0.308$, $\Omega_{\Lambda} = 0.692$ and $h = 0.678$.

\section{Modelling the IGM damping wing}
\label{section:models}

\subsection{The Sherwood-Relics simulations}

To generate mock IGM damping wings, we analyse simulations from the Sherwood-Relics simulation suite\footnote{\url{https://www.nottingham.ac.uk/astronomy/sherwood-relics}} \citep{puchwein2023}. These are a set of cosmological hydrodynamical simulations that build upon the original Sherwood suite \citep{bolton2017} and which were designed to model the evolution of the IGM during and after the epoch of reionization. The simulations were performed with a modified version of the smoothed particle hydrodynamical code \textsc{p-gadget-3} \citep{springel2005}. The fiducial simulation we analyse has a box size of $40 \, h^{-1}$ cMpc and $2 \times 2048^3$ gas and dark matter particles, resulting in particle masses of $M_{\rm gas} = 9.97 \times 10^{4} \, h^{-1} \, M_{\odot}$ and $M_{\rm dm} = 5.37 \times 10^{5} \, h^{-1} \, M_{\odot}$. The gravitational softening used was $l_{\rm soft} = 0.78 \, h^{-1}$ kpc. We further analyse a larger, lower resolution volume with box size $160 \, h^{-1}$ cMpc and $2 \times 2048^3$ gas and dark matter particles. This simulation has $M_{\rm gas} = 6.38 \times 10^{6} \, h^{-1} \, M_{\odot}$, $M_{\rm dm} = 3.44 \times 10^{7} \, h^{-1} \, M_{\odot}$ and $l_{\rm soft} = 3.13 \, h^{-1}$ kpc. Star formation in these simulations is treated using a computationally efficient sub-grid model, where any gas with temperature $T < 10^{5}$ K and density 1000 times the cosmic mean density is immediately converted into star particles. Although this approach is highly simplified, it should not affect the properties of the low-density IGM in which we are interested \citep{viel2004}.
 
To account for the patchy nature of reionization, we utilise a subset of the Sherwood-Relics simulations that were performed with an inhomogeneous UV background. Although it is now becoming feasible to perform fully-coupled radiation-hydrodynamic galaxy formation simulations of cosmic reionization \citep[see][for a recent review]{gnedin2022}, a different approach has been taken in the Sherwood-Relics project. We give only a brief overview of the method here, but note that it is described in greater detail in \citet{puchwein2023}. We first perform a cosmological simulation with \textsc{p-gadget-3} using the uniform UV background of \citet{puchwein2019}, saving snapshots of this simulation every 40 Myr. We then map these snapshots onto $2048^3$ Cartesian grids and post-process them with the GPU-powered radiative transfer code \textsc{aton} \citep{aubert2008, aubert2010}. Multiple iterations of the radiative transfer simulations are performed, varying the redshift evolution of the volume emissivity assumed in the simulation until the desired reionization history is reached. We use a simple model for the sources, such that their emissivity is proportional to their halo mass \citep{iliev2006,chardin2015}. For each of the $2048^3$ cells on the grid, the reionization redshift and redshift evolution of the amplitude of the UV background are calculated from the outputs of the radiative transfer code. These data are then used as inputs to a new \textsc{p-gadget-3} simulation, to capture the effect of a spatially fluctuating UV background due to patchy reionization in the hydrodynamic simulation.

The advantages to this hybrid approach are that it is computationally much cheaper than performing a full radiation-hydrodynamic simulation, while still allowing us to track the hydrodynamic response of the gas to reionization. We can also guarantee the reionization history of our final patchy simulation, which can be difficult to calibrate in self-consistent galaxy formation simulations of reionization. There are of course also some disadvantages to this method, such as the finite spatial and temporal resolution of our ionizing background, but we do not expect this to have a significant effect on the diffuse intergalactic gas that is our primary focus here.

\begin{figure}
\includegraphics[width=\columnwidth]{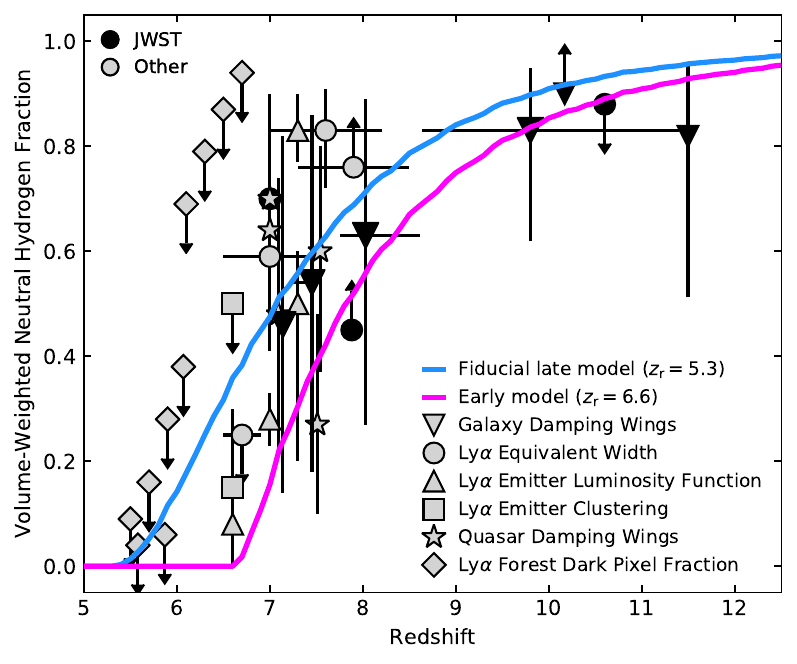}
    \caption{Comparison of the reionization histories analysed in this paper to a selection of constraints from the literature on the redshift evolution of the volume-averaged neutral fraction of the IGM. The solid blue line shows our fiducial model, where reionization ends at $z=5.3$. The solid pink line shows a more extreme early model, where reionization ends at $z=6.6$. For comparison, we show reionization constraints from observations of galaxy damping wings \citep{curtislake2023, hsiao2023, umeda2023}, Ly$\alpha$ equivalent widths \citep{mason2018, mason2019, bolan2022, bruton2023, jones2023, morishita2023}, Ly$\alpha$ emitter luminosity functions \citep{inoue2018, morales2021}, Ly$\alpha$ emitter clustering \citep{sobacchi2015, ouchi2018}, quasar damping wings \citep{davies2018,wang2020,greig2022} and dark pixel fractions of the Ly$\alpha$ forest \citep{mcgreer2015, jin2023}. The colour of the points highlights the spectacular redshift reach of JWST (black) compared to other sources (grey).} 
\label{fig:HI_history}
\end{figure}

Our fiducial reionization history is shown in Figure \ref{fig:HI_history} (blue line) and compared against different probes of the IGM neutral fraction. Our preferred reionization history is a model where reionization ends at $z=5.3$ (where we define the ``end'' of reionization as the redshift where the IGM first becomes 99.9 per cent ionized by volume). This is in agreement with the tight constraints on the end of reionization obtained from the opacity of the Ly$\alpha$ forest \citep{bosman2022}. This model is 50 per cent ionized at $z=7.1$, which sits within the scatter of the various probes of the IGM neutral fraction at that redshift. To explore the effect of changing the reionization history on our simulated damping wings, we also analyse an alternative reionization history (shown by the pink line in Figure \ref{fig:HI_history}). In this early reionization model, the IGM is 99.9 per cent ionized at $z=6.6$ and 50 per cent ionized at $z=7.8$. As this early model is in tension with many of the constraints shown in Figure \ref{fig:HI_history}, we present it only as an example of how our results would change in a more extreme scenario, and do not advocate it as a preferred reionization history.

\subsection{Simulated absorption spectra}
\label{section:describe_spectra}

\begin{figure*}
	\includegraphics[width=1.8\columnwidth]{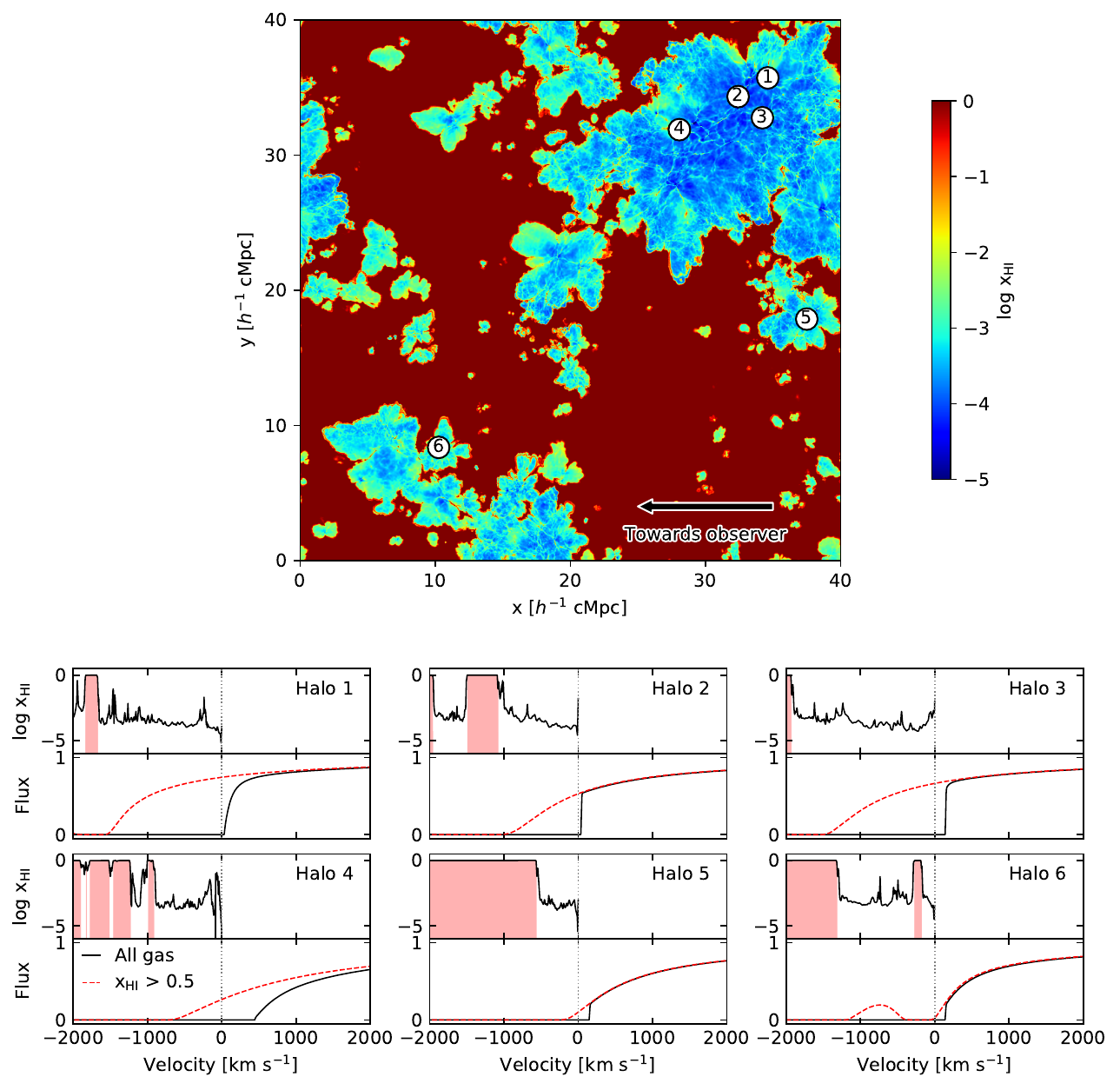}
    \caption{Examples demonstrating the diversity of our simulated IGM damping wings at $z=8$. The map on the top shows the neutral fraction of the gas in a 19.53 $h^{-1}$ ckpc slice through our simulation. The white circles mark the locations of the six haloes plotted below, with the numbers identifying the different haloes, connecting them to the large-scale ionization field of our reionization simulations.  The arrow in the bottom right shows the direction along which the spectra were calculated. The six subplots on the bottom two rows show the results from lines of sight through different haloes. The top panel of each subplot shows the neutral gas fraction along the line of sight. The regions intersecting islands of neutral gas (which we define as having a neutral fraction $x_{\rm HI} > 0.5$) are highlighted in red. The bottom panel of each subplot shows the corresponding Ly$\alpha$ absorption spectrum. The solid black lines show the spectra calculated taking the contribution of all gas (both in ionized bubbles and neutral islands) into account. The dashed red line shows the spectra calculated assuming the gas in the bubbles is completely ionized, and only the neutral islands contribute to the Ly$\alpha$ optical depth.}
    \label{fig:HI_map}
\end{figure*}

To construct our mock damping wings, we extract lines of sight from simulation snapshots at $z=7,8,9$ and $10$. As we are interested in the IGM absorption arising in gas in the foreground of galaxies, we take these lines of sight through the 100 most massive haloes at each redshift, extracting the gas neutral hydrogen density, temperature and peculiar velocity. We orient these lines of sight in the $\pm x, \pm y$ and $\pm z$ directions around the halo and continue them for 20 $h^{-1}$ cMpc from the halo centre. This results in 600 sightlines in total. The width of each pixel in the sightline is 19.53 $h^{-1}$ ckpc for our fiducial 40 $h^{-1}$ cMpc $2048^3$-particle simulation volume and 78.13 $h^{-1}$ ckpc for our larger, lower resolution 160 $h^{-1}$ cMpc $2048^3$-particle volume.

As neutral gas at large distances can also contribute to the strength of the IGM damping wing, we stitch on five additional random sightlines length 40 $h^{-1}$ cMpc through the simulation volume. As we would like to account for the redshift evolution of the IGM along the line of sight, we make use of the on-the-fly sightlines that were extracted from the Sherwood-Relics simulation at redshift frequency $\Delta z = 0.1$. Using these data, we stitch on subsequent sightlines taken from increasingly lower redshifts, consistent with the redshift interval corresponding to the comoving distance between the halo position and the beginning of the new sightline. This results in final lines of sight that are each 220 $h^{-1}$ cMpc long.

From these sightlines, we then sum up the contributions of the gas in front of the halo to the Ly$\alpha$ optical depth using the analytic approximation to the Voigt profile presented in \citet{teppergarcia2006}. Due to the simplified star formation prescription used in our simulations, we do not expect to recover the distribution of neutral gas within the circumgalactic medium of the halo. We further aim to model only the contribution of the IGM to the damping wing. We therefore exclude the contribution of any gas within the virial radius of the halo by default \citep[see also][]{weinberger2018}. As the redshift of the observed galaxies is known from their emission lines, we renormalise the gas peculiar velocity along the line of sight such that the halo has a velocity of $0 \, \rm{km}\,\rm{s}^{-1}$. As the damping wing extends redward of the Ly$\alpha$ source frame redshift, we also record the optical depth in the mock spectra redwards of the halo location up to a velocity corresponding to 200 $h^{-1}$ cMpc ``behind'' the halo, but of course only accounting for absorption by gas that is in front of the halo in position space. Examples of our simulated IGM damping wings are shown in Figure \ref{fig:HI_map}.

\subsection{The Miralda-Escudé model for the IGM damping wing} 
\label{section:ME98_model}

\begin{figure*}
\includegraphics[width=2\columnwidth]{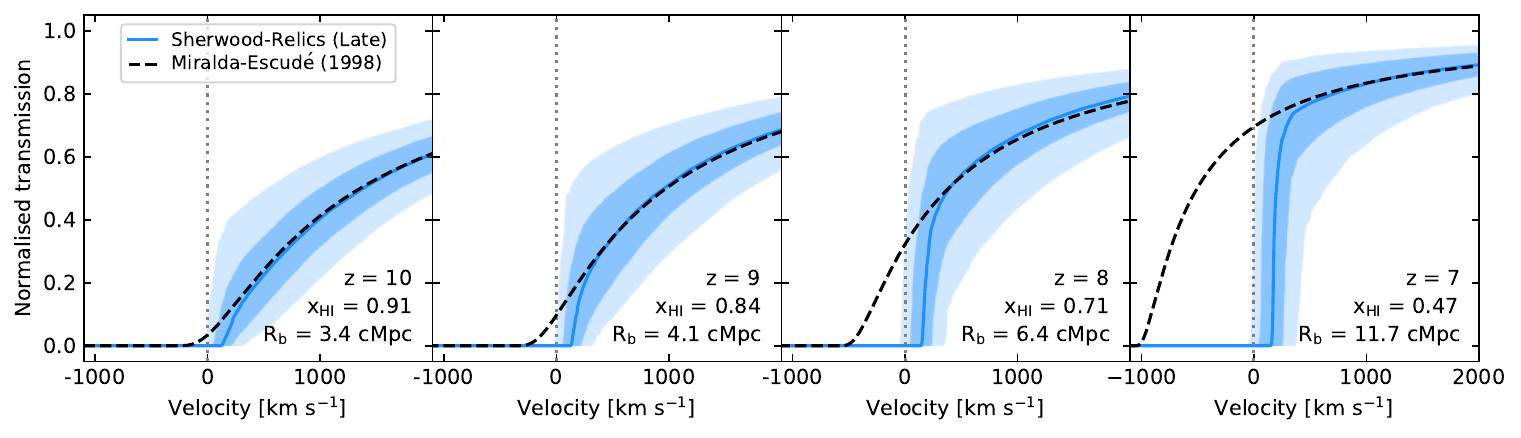}
    \caption{Comparison of the simulated Sherwood-Relics damping wings for the late reionization history with the \citet{miraldaescude1998} model. The different panels show results at different redshifts. The blue solid line shows the median Ly$\alpha$ transmission for the simulated sightlines as a function of velocity. The light and dark shaded regions encompass the 68 and 95 per cent scatter of the transmission, respectively. The dashed black line is the \citet{miraldaescude1998} model computed assuming the volume-weighted \ion{H}{i} fraction and the median bubble size in the simulation as indicated on each panel.}   
   \label{fig:compare_model}
\end{figure*}

As the goal of this work is to contrast the IGM damping wings constructed from cosmological simulations with the analytic \citet{miraldaescude1998} model, we recap the details of that model here for completeness. 

It is useful to first define the \citet{gunn1965} optical depth for a uniform IGM at redshift $z$, 
\begin{align}
\label{eqn:GP}
\tau_{\rm GP}(z) &= \frac{3\lambda_\alpha^3\Lambda_\alpha n_{\rm HI}}{8\pi H(z)},\\
 &\simeq 5.62\times 10^{5} x_{\rm HI} \left(\frac{\Omega_{\rm b}h^{2}}{0.022}\right)\left(\frac{\Omega_{\rm m}h^{2}}{0.142}\right)^{-1/2}\left( \frac{1+z}{9}\right)^{3/2},
\end{align}
\noindent
where $\Lambda_{\alpha}$ is the Ly$\alpha$ decay constant, $\lambda_{\alpha}$ is the Ly$\alpha$ rest frame wavelength and $H(z)$ is the Hubble parameter. At high redshift, this can be approximated as $H(z) \simeq H_0 \Omega_{\rm m}^{1/2} (1+z)^{3/2}$. The background neutral hydrogen density is defined as $n_{\rm HI} = x_{\rm HI}\langle n_{\rm H} \rangle $, where $\langle n_{\rm H}\rangle =\rho_{\rm crit}\Omega_{\rm b}(1-Y)(1+z)^{3}/m_{\rm H}$. Here, $\rho_{\rm crit}$ is the critical density at $z=0$, $\Omega_{\rm b}$ is the baryon density, $Y$ is the helium mass fraction and $m_{\rm H}$ is the mass of a hydrogen atom.

Using Equation \ref{eqn:GP}, we can then define the optical depth of the IGM damping wing,
\begin{equation}
\label{eqn:DW}
\tau_{\rm D}(z) = \frac{\tau_{\rm GP}(z_{\rm s})R_{\alpha}}{\pi}\left(\frac{1+z}{1+z_{\rm s}}\right)^{3/2} \left[I\left(\frac{1+z_{\rm b}}{1+z}\right)-I\left(\frac{1+z_{\rm n}}{1+z}\right)\right], 
\end{equation}
\noindent
where $\tau_{\rm D}$ is the optical depth along the line of sight evaluated at redshift $z$, $\tau_{\rm GP}$ is the Gunn-Peterson optical depth defined above, $z_{\rm s}$ is the redshift of the source, $z_{\rm b}$ is the redshift of the edge of the ionized bubble and $z_{\rm n}$ is the redshift where reionization is defined to end. $R_{\alpha}$ is defined as $\Lambda_{\alpha}\lambda_{\alpha}/(4\pi c)$, where $c$ is the speed of light. Finally, the function $I(x)$ is given by
\begin{align}
    I(x)&=\frac{x^{9/2}}{1-x}+\frac{9}{7}x^{7/2}+\frac{9}{5}x^{5/2}+3x^{3/2} +9x^{1/2} \nonumber \\
    &-\frac{9}{2}\ln\left({\frac{1+x^{1/2}}{1-x^{1/2}}}\right). 
\end{align}

\noindent
Note this function is only well defined for $x=(1+z_{\rm b})/(1+z) < 1$. We therefore only compute the optical depth of the damping wing for redshifts $z > z_{\rm b}$, and at lower redshifts set the IGM optical depth to $\tau_{\rm GP}(z)$.

\section{Comparison of IGM damping wing models}
\label{section:compare_model}

\subsection{Comparison with late reionization model}
\label{section:compare_late}

We show the median and scatter of the IGM damping wings computed from Sherwood-Relics with the fiducial late reionization history with the blue lines and shaded regions in Figure \ref{fig:compare_model}. Each panel corresponds to a different redshift. We recover the expected result that as we move to higher redshifts and further into the epoch of reionization, the IGM damping wings become stronger due to the increasing volume-weighted average neutral fraction of the IGM.

Figure \ref{fig:compare_model} further shows the results of our comparison between the Sherwood-Relics late reionization model and the \citet{miraldaescude1998} analytic model (denoted by the dashed black line). To make a fair comparison between the two models, we compute the \citet{miraldaescude1998} model assuming parameters from our simulation. We measure the volume-weighted average neutral fraction from the simulation output at the redshift of the halo. Note that this will not take any evolution of the IGM along the line of sight into account, which is taken into account in an approximate manner in our simulated sightlines. We further assume the end of reionization to be the redshift where the IGM is 99.9 per cent ionized by volume. For the size of the ionized bubble, we measure the distance between the position of the halo where we begin our sightlines and the point where the neutral fraction of the gas first exceeds $x_{\rm HI} = 0.5$ along each line of sight. We then use the redshift corresponding to the median of these bubble sizes along all lines of sight as input to Equation \ref{eqn:DW}. The values used are indicated on the bottom right of each panel of Figure \ref{fig:compare_model}.

 We find that while the \citet{miraldaescude1998} model does an excellent job of recovering the median IGM damping wing predicted by the Sherwood-Relics simulation at velocities a few 100 $\rm{km} \, \rm{s}^{-1}$ redward of the systemic velocity, it overpredicts the IGM transmission otherwise. The deviation between the analytic model and the simulations begins slightly redward of the systemic velocity and grows towards bluer wavelengths. The discrepancy is small at higher redshift when the ionized bubble sizes are small and the volume-weighted neutral fraction of the IGM is large, but becomes increasingly apparent towards lower redshift. For example, the \citet{miraldaescude1998} model at $z=7$ calculated using the parameters from our simulation predicts that the IGM transmission should be 50 per cent at a velocity of $-500 \, \rm{km} \, \rm{s}^{-1}$, while in Sherwood-Relics no transmitted flux is expected at all at that velocity.

\subsection{Comparison with additional reionization models} 

\begin{figure*}
\includegraphics[width=2\columnwidth]{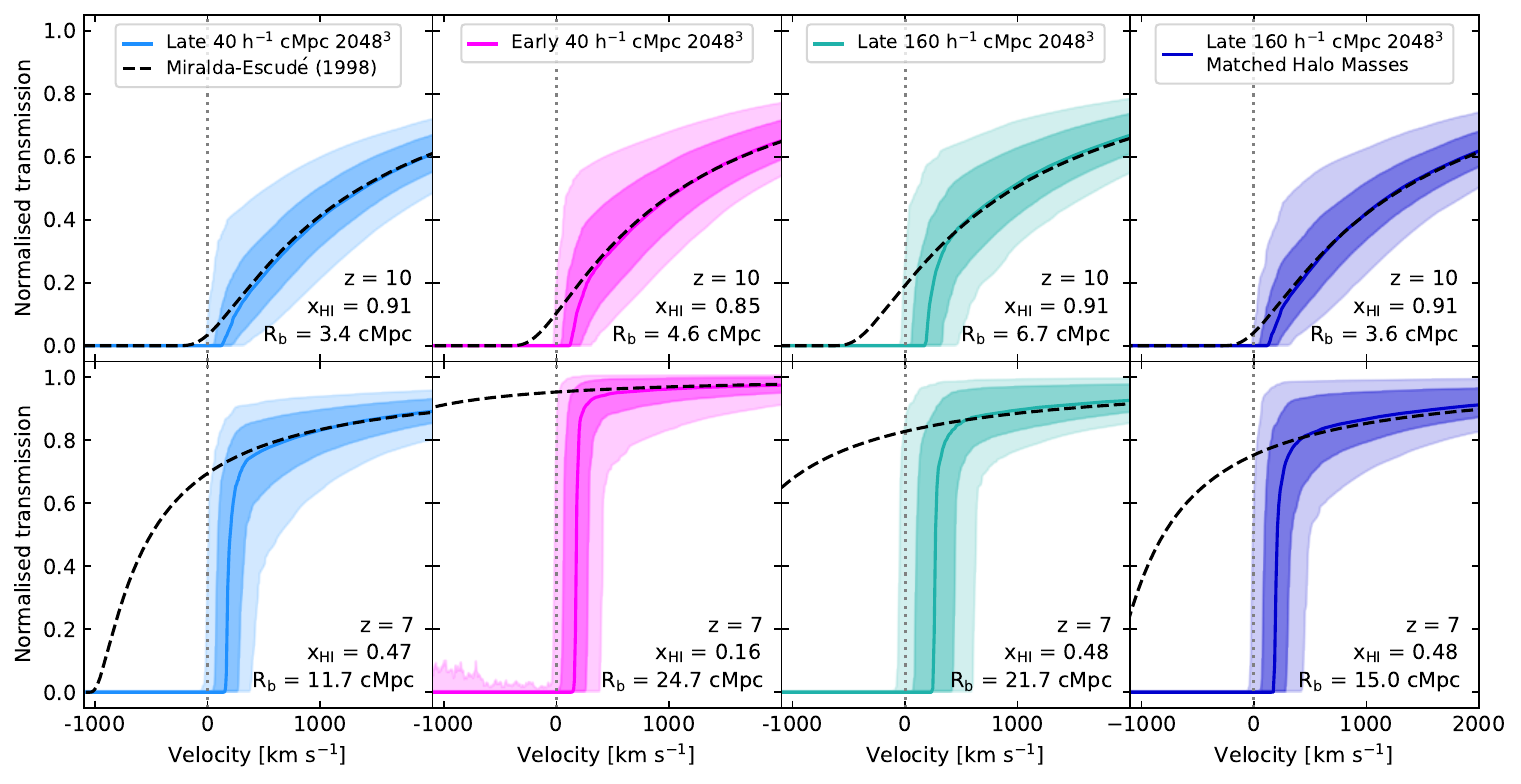}
    \caption{Comparison of the simulated damping wings from different simulations of inhomogeneous reionization from the Sherwood-Relics suite with the \citet{miraldaescude1998} model. The different rows show results at different redshifts (\textit{top:} $z=10$, \textit{bottom:} $z=7$).  \textit{First column:} Absorption spectra as shown in Figure \ref{fig:compare_model}, calculated from the fiducial late reionization simulation in the high resolution 40 $h^{-1}$ cMpc volume. \textit{Second column:} Absorption spectra calculated from the early reionization simulation in the high resolution 40 $h^{-1}$ cMpc volume. \textit{Third column:} calculated from the fiducial late reionization simulation in the lower resolution 160 $h^{-1}$ cMpc volume. \textit{Fourth column:} Absorption spectra calculated from the fiducial late reionization simulation in the lower resolution 160 $h^{-1}$ cMpc volume for a similar halo mass range as in the 40 $h^{-1}$ cMpc volume.} 
   \label{fig:compare_model_history_boxsize}
\end{figure*}

We next explore the IGM damping wings generated from different patchy reionization simulations from the Sherwood-Relics simulation suite. The results are shown in Figure \ref{fig:compare_model_history_boxsize} for simulation snapshots at redshifts $z=7$ and $10$. In each case, we overplot the results from the \citet{miraldaescude1998} damping wing model, calculated from the properties of the corresponding simulation. For easy comparison, the first column repeats the results from the fiducial late reionization model as already discussed in Section \ref{section:compare_late} and shown in Figure \ref{fig:compare_model}. The second column contrasts this with simulated IGM damping wings computed from the early reionization model. At fixed redshift, this early model has both lower volume-weighted average neutral fractions and larger bubble sizes than the late model, as indicated in the lower right of each panel of Figure \ref{fig:compare_model_history_boxsize}. This results in weaker IGM damping wings in the early reioniziation model than in the late reionization model at a given redshift. The other notable difference is that at $z=7$ in the early reionization model, there is some transmitted flux visible in the Ly$\alpha$ forest blueward of the systemic redshift. However, even in this extreme reionization model, this flux is still much lower than the corresponding prediction for the IGM transmission of the \citet{miraldaescude1998} model.
 
We next explore how our results depend on the volume of our reionization simulation, as it has been shown that volumes of several hundred cMpc are required to fully capture the patchiness of reionization \citep{iliev2014}, in contrast to the fiducial 40 $h^{-1}$ cMpc volume we analyse here. The third column of Figure \ref{fig:compare_model_history_boxsize} shows the simulated IGM damping wings computed from a 160 $h^{-1}$ cMpc simulation volume, with a near-identical reionization history to the late reionization 40 $h^{-1}$ cMpc volume plotted in the first column over the redshift range probed here. We find that at a given redshift, although the volume-weighted average neutral gas fraction is the same in the two simulations, the bubbles are larger in the larger volume by approximately a factor of two. The result of this is that the sightlines generated from the larger volume show more transmission in the IGM damping wing profile at lower wavelengths, with even a small amount of transmission occurring just blueward of the systemic velocity. However, this effect is somewhat offset by the larger infall velocities associated with the more massive halo masses that are found in the larger volume. At $z=7$, the median mass of the 100 most massive haloes in the 40 $h^{-1}$ cMpc volume is $9.0 \times 10^{10} h^{-1} M_{\odot}$, compared with $4.3 \times 10^{11} h^{-1} M_{\odot}$ in the 160  $h^{-1}$ cMpc volume. This results in a larger offset between the systemic velocity and the median of the sharp cutoff in the Ly$\alpha$ transmission in the larger volume than in the smaller volume.

To investigate how the two simulations compare for a more similar population of haloes, we take a different set of lines of sight from the 160  $h^{-1}$ cMpc volume. This time we select a sample of haloes that have a distribution of masses similar to the 100 most massive haloes in the 40 $h^{-1}$ cMpc volume. The results of this test are shown in the fourth column of Figure \ref{fig:compare_model_history_boxsize}. We find that the median Ly$\alpha$ transmission profile now looks more similar in the 160 $h^{-1}$ cMpc and 40 $h^{-1}$ cMpc volumes, as
the velocity offset is now much closer between the two models. We also find a median bubble size that is smaller around the lower mass haloes. However, the scatter in bubble sizes is quite different, with the 68 (95) per cent range in bubble size a factor of two (three) larger in the larger volume simulation, most likely because these lower mass haloes are clustered around more massive haloes that can carve out larger ionized bubbles. The result of this is slightly weaker IGM damping wings in the 160 $h^{-1}$ cMpc compared to the 40 $h^{-1}$ cMpc volume, even when probing haloes of the same mass at fixed volume-weighted average neutral fraction.

However, in all cases, we still recover the trend that our simulated IGM damping wings are not described well by the \citet{miraldaescude1998} analytic model blueward of the systemic redshift, independent of the reionization history, simulation volume or host halo mass.

\subsection{Reasons for difference between analytic and simulated damping wing models}
\label{section:model_diff}

\begin{figure*}
\includegraphics[width=2\columnwidth]{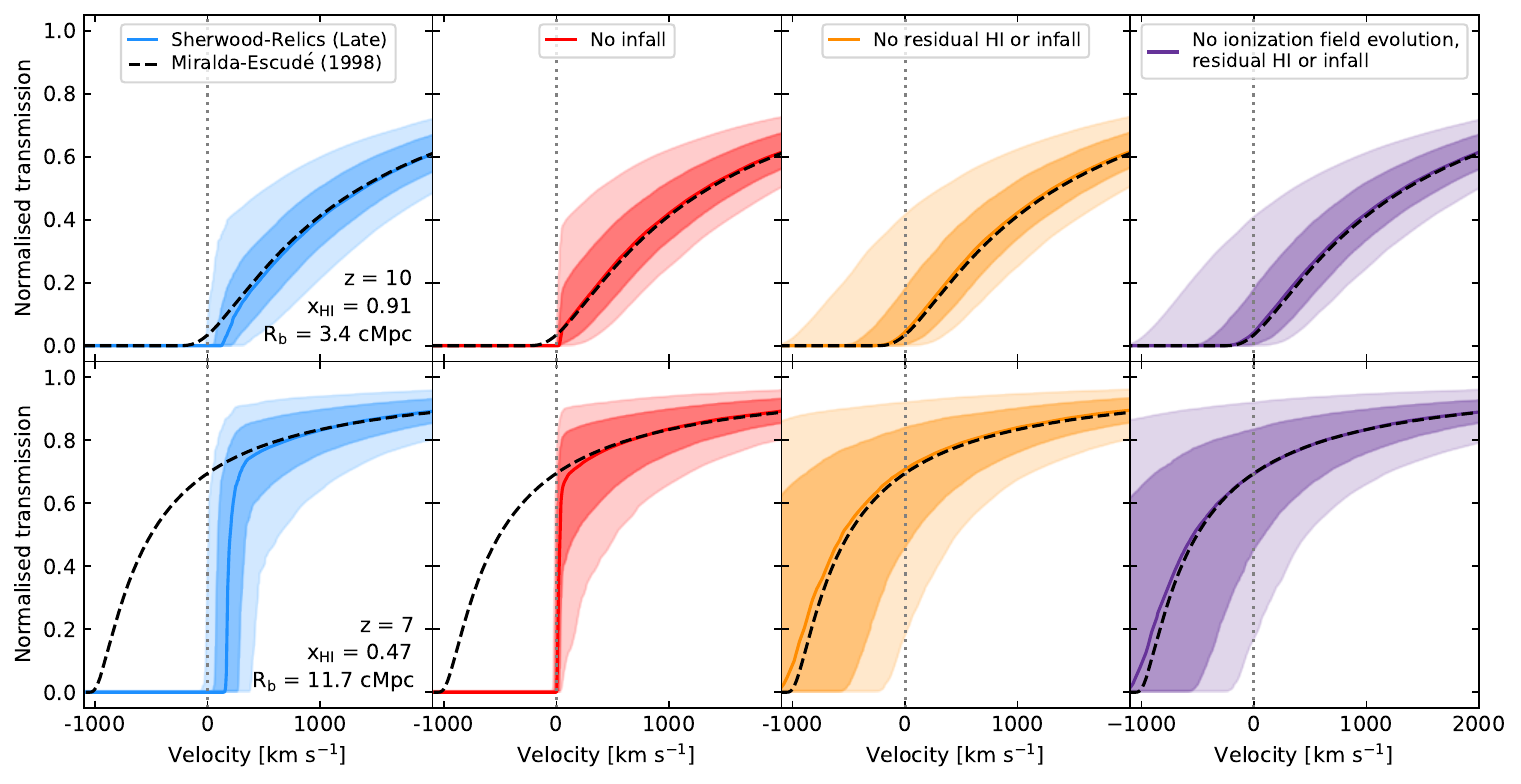}
    \caption{Comparison of the simulated Sherwood-Relics damping wings with the \citet{miraldaescude1998} model. The different rows show results at different redshifts (\textit{top:} $z=10$, \textit{bottom:} $z=7$). The different columns compute the absorption spectra with and without different physical effects. \textit{First column:} Absorption spectra as shown in Figure \ref{fig:compare_model}, calculated with all the relevant physical effects. \textit{Second column:} Absorption spectra calculated without the effect of gas peculiar velocities. \textit{Third column:} Absorption spectra calculated without any residual neutral hydrogen inside the ionized bubbles and neglecting the gas peculiar velocities. \textit{Fourth column:} Absorption spectra calculated without accounting for any redshift evolution of the ionization state along the line of sight, without any residual neutral hydrogen inside the ionized bubbles and neglecting the gas peculiar velocities.}
   \label{fig:compare_model_vpec_ion}
\end{figure*}

We investigate the root of the difference between the IGM damping wings from the Sherwood-Relics simulations and the \citet{miraldaescude1998} model in Figure \ref{fig:compare_model_vpec_ion}. The first column of this figure shows the IGM damping wings at $z=10$ and $z=7$. These are exactly the same as displayed in Figure \ref{fig:compare_model}, and are calculated as described in Section \ref{section:describe_spectra}. The second column shows the IGM damping wings calculated for the same sightlines, but now we compute the Ly$\alpha$ optical depth without taking the peculiar velocity of the gas into account. The effect of this is to produce a sharp cutoff in the IGM transmission at a velocity of $0 \, \rm{km} \, \rm{s}^{-1}$, whereas previously this cutoff occurred at a range of different velocities redward of the systemic velocity, due to the range of motions of the local gas falling into the host halo. By disregarding this infalling gas and shifting the cutoff in the IGM transmission to bluer wavelengths, we find that we improve the agreement between the Sherwood-Relics damping wings and the \citet{miraldaescude1998} damping wing model such that there is near perfect agreement for all transmission redward of the halo redshift. However, we still find conflicting predictions for the level of transmission expected blueward of the halo redshift. 

Next, we investigate the impact of residual neutral gas in the ionized bubbles. We recompute the Ly$\alpha$ optical depth along our lines of sight, but now assume that all of the gas within the ionized bubble surrounding the host halo is completely ionized. We achieve this by finding the first neutral island along our line of sight (which we define as a pixel with $x_{\rm HI} > 0.5$), and then setting the neutral fraction in all pixels between the halo and that first neutral island to have $x_{\rm HI} = 0$. The reason that we do not just remove the residual neutral gas in all bubbles, rather than just the host bubble, is that this will produce additional transmission along the line of sight. See, for example, the additional transmission peak separated from the IGM damping wing in Halo 6 of Figure \ref{fig:HI_map}. This would confuse our damping wing signal when we stack our absorption spectra, and hence our comparison with the \citet{miraldaescude1998} model. We therefore chose to isolate the effect of the residual neutral gas only in the vicinity of the host halo.

We show the median transmission profiles and their scatter in the third column of Figure \ref{fig:compare_model_vpec_ion}, where we have now excluded both the residual neutral gas and neglected the peculiar velocities of the gas. We find that this makes a large difference to the predicted transmission blueward of the host halo redshift, and there is now very good agreement between the damping wings calculated from the Sherwood-Relics simulation and the \citet{miraldaescude1998} model. As has previously been noted in other works, residual neutral gas inside the ionized bubbles should play a large role (see, e.g., \citealt{MesingerHaiman2004} and \citealt{bolton2007} for a discussion in the context of quasar proximity zones, and \citealt{mason2020} in the context of Ly$\alpha$ emitting galaxies). Although the gas in the ionized bubbles in the Sherwood-Relics simulation is highly ionized, it still has a neutral fraction of order $x_{\rm HI} \sim 10^{-3}$ (see Figure \ref{fig:HI_map}). From Equation \ref{eqn:GP}, this will result in a Ly$\alpha$ optical depth $\tau_{\rm Ly\alpha} \sim 500$, more than enough to completely saturate the IGM absorption.

After removing both the effects of peculiar velocities and residual neutral gas, we find much better agreement between our simulated IGM damping wings and the \citet{miraldaescude1998} model. There still remains a small difference between the analytic model and the median IGM damping wing predicted from our simulations, with the simulations predicting slightly more transmission at velocities blueward of the host halo redshift. As a final test, we therefore investigate the effect of neglecting evolution of the ionization state of the gas along the line of sight. As described in Section \ref{section:describe_spectra}, we account for this in an approximate way by stitching together lines of sight from simulation outputs at different redshifts. We check what effect this has by recomputing the lines of sight, but this time now stitching together lines of sight from a fixed redshift, such that there is no evolution in the average ionization field across the 220 $h^{-1}$ cMpc line of sight, although individual pixels will still be ionized or neutral depending on their location within the simulation volume. We find however that neglecting the evolution of the IGM along the line of sight has only a small effect on our simulated damping wings.

In summary, we find that the most significant reason for the difference between the predictions for the IGM damping wing from the Sherwood-Relics simulations and the \citet{miraldaescude1998} model is the residual neutral gas within the ionized bubbles. While the effects of peculiar motions of infalling gas and evolution of the IGM along the line of sight towards the observer also play a small role, these are subdominant to the resonant absorption caused by the gas within the bubbles.

\section{Comparison of models and observations}
\label{section:compare_observations}

We next seek to make contact between the IGM damping wings constructed from our simulations of inhomogeneous reionization, and the the first observations of galaxy damping wings performed with JWST. In particular, we investigate the recently inferred 100 cMpc-sized ionized bubbles during reionization \citep{umeda2023} and the observations of proximate DLAs in high-redshift galaxies \citep{heintz2023}.

\subsection{Large ionized bubbles during reionization}
\label{sec:compare_umeda23}

\begin{figure*}
\includegraphics[width=2\columnwidth]{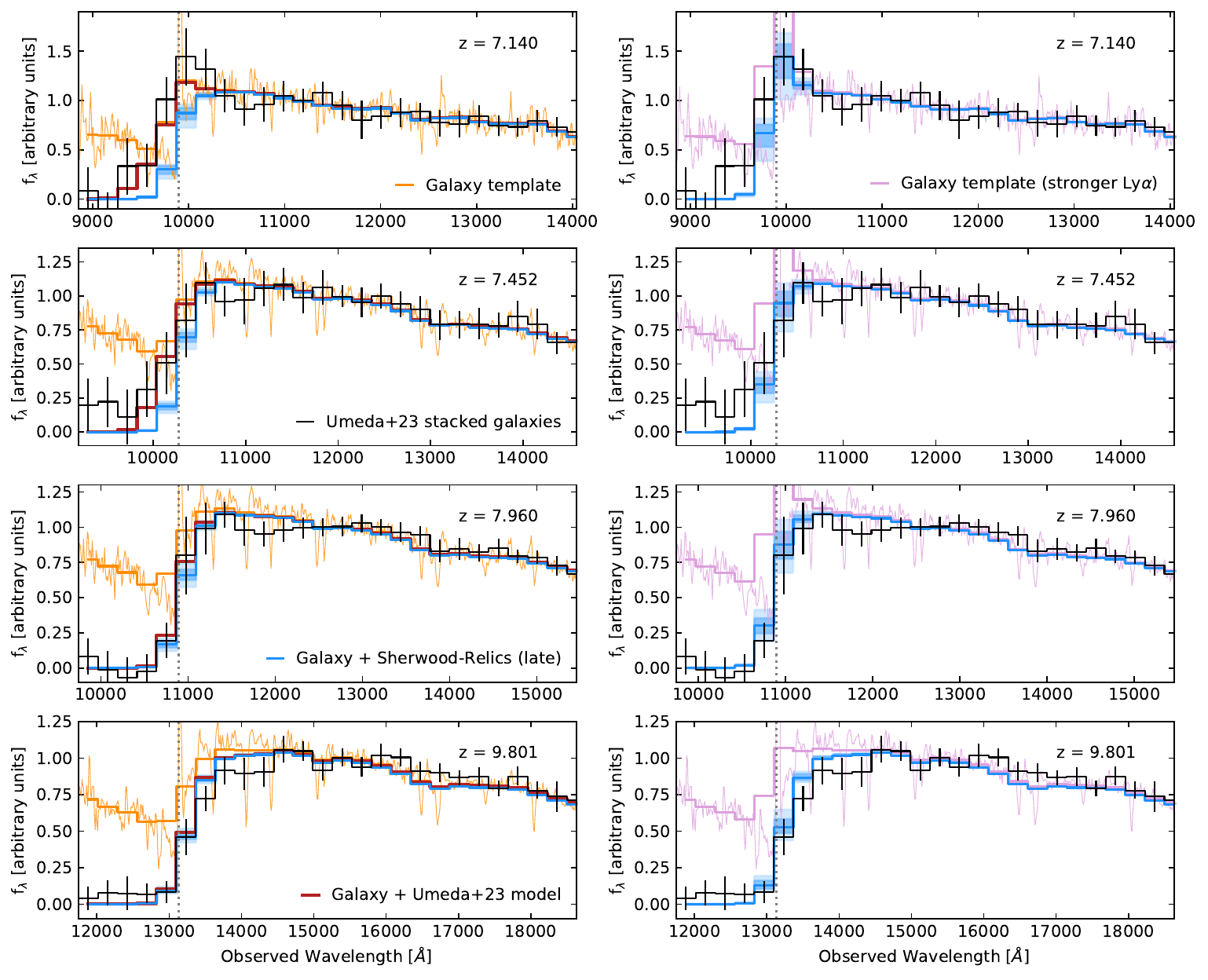}
    \caption{\textit{Left column:} Comparison of modelled IGM damping wings and stacks of galaxy spectra from \citet{umeda2023}. The stacked observed galaxy spectra are shown in black. Each row corresponds to a different redshift, at $z=7.140, 7.452, 7.960$ and $z=9.801$. The thin orange lines show stacked spectra at lower redshift from the VANDELS survey \citep{cullen2019}, rescaled to match the continua of the higher redshift galaxies. The thick orange lines show the same spectra but after convolution and binning to coarser pixels. The thick red line shows the template intrinsic galaxy spectra convolved with the best-fitting IGM damping wing model of \citet{umeda2023}. The blue lines and shaded region show the intrinsic galaxy spectra convolved with absorption spectra generated from the Sherwood-Relics (late) simulation. \textit{Right column:} As before, but now using a template spectrum where we have changed the strength of the Ly$\alpha$ emission line by hand, such that it is now increased in amplitude by a factor of three.}
   \label{fig:umeda23_stacks}
\end{figure*}

We first turn our attention to the IGM damping wings measured from stacks of galaxies presented in \citet{umeda2023}. The sample consists of spectra of 26 galaxies at $7 <z < 12$ obtained from multiple JWST programs \citep{arrabalharo2023, hsiao2023, finkelstein2023}. These galaxies were divided into four redshift bins with median redshifts $z=7.140, 7.452, 7.960$ and $9.801$. To see the imprint of the IGM damping wing, it was necessary to compare the observed spectrum with a model for the intrinsic spectrum. To properly account for absorption in the circumgalactic medium of the galaxies and isolate the effect of the IGM, \citet{umeda2023} used stacks of galaxies in bins of stellar mass at $2.5 < z < 5$ from the VANDELS survey \citep{cullen2019}. By choosing the template in the stellar mass bin that best matched the observed galaxy continuum far redward of the Ly$\alpha$ source frame wavelength, and finding the best-fitting \citet{miraldaescude1998} damping wing model in the wavelength region close to Ly$\alpha$, they obtained constraints on the redshift evolution of both the volume-weighted average neutral fraction of the IGM and the sizes of the ionized bubbles.

As we wish to compare the IGM damping wings generated from the Sherwood-Relics simulations, we first repeat the process to generate the template intrinsic spectra. We use the stacked spectra from the stellar mass bins that \citet{umeda2023} found to best fit their stacked spectra. Namely, we use the $\log(M_{\star}/M_{\odot}) = 8.16-8.70$ stack at $z=7.140$, the 
$\log(M_{\star}/M_{\odot}) = 8.70-9.20$ stack at $z=7.452$ and $7.960$ and the $\log(M_{\star}/M_{\odot}) = 9.50-9.65$  stack at  $z=9.801$. We renormalise these spectra by a factor $A (\lambda/1800 \mbox{\normalfont\AA})^b$, where $A$ and $b$ are parameters that can be varied until the $\chi^2$ between the template stacks and the observed stacks is minimised. Following \citet{jones2023}, to account for the low spectral resolution of the NIRSpec PRISM mode\footnote{We took the resolving power as a function of wavelength from \url{https://jwst-docs.stsci.edu/jwst-near-infrared-spectrograph/nirspec-instrumentation/nirspec-dispersers-and-filters}}, we convolve the spectra with a Gaussian of an appropriate width. We further bin the template spectra to pixels of 25 \AA \, in the rest frame, to match the observed stacks. As in \citet{umeda2023}, we perform this fitting the wavelength range between the wavelength where an IGM damping wing generated with the \citet{miraldaescude1998} model for a neutral IGM with no ionized bubble first reaches a transmission of 90 per cent at the blue end, and 2200 \AA \, rest frame wavelength at the red end. The resulting template spectra are shown in orange in the left panel of Figure \ref{fig:umeda23_stacks} (the thin lines are before convolution and binning, and the thick lines are afterwards) and can be compared with the stacked observed spectra from \citet{umeda2023} shown in black. Having generated the template intrinsic spectra, we then convolve these with models for the IGM damping wing. We first use the best-fitting parameters measured in \citet{umeda2023}, which are volume-weighted average neutral fractions $\langle x_{\rm HI} \rangle_{\rm v} = (0.46, 0.54, 0.63, 0.83)$ and bubble sizes $R_{\rm b} = (149, 96.1, 16.5 , 5.04)$ cMpc at redshift $z = (7.140, 7.452, 7.960, 9.801)$. We show the resulting curves generated from the galaxy template, convolved with these damping wing models, as the dark red curves in Figure \ref{fig:umeda23_stacks}.

We next repeat this process, but instead now use the simulated IGM damping wings generated from the Sherwood-Relics late reionization simulation. We generate spectra from the snapshots at $z=7,7.5,8$ and $10$, but rescale the densities and velocities to match the median redshift in each bin of the \citet{umeda2023} sample. The resulting galaxy templates convolved with the simulated damping wings are shown by the blue curves and light (dark) shaded regions in Figure \ref{fig:umeda23_stacks}, which represent the median and 68 (95) per cent scatter of the distribution. We find that in the two highest redshift bins, at $z=9.801$ and 7.960, there is good agreement between the predictions from the best fit IGM damping wing from \citet{umeda2023} and the Sherwood-Relics simulations. We do find that the Sherwood-Relics simulation seems to underpredict the strength of the damping wing at $z=9.801$, but this is also seen in the \citet{umeda2023} IGM damping wing model, and may be due to the influence of proximate absorbers as discussed in Section \ref{sec:compare_DLAs}.

We find, however, that there is a significant difference in the expected transmitted flux between the Sherwood-Relics damping wings and the best fit \citet{umeda2023} IGM damping wing model at $z=7.140$ and 7.452. The intrinsic galaxy template convolved with the Sherwood-Relics model produces too little transmitted flux in the pixels close to the Ly$\alpha$ source frame wavelength. At first glance, this suggests that the damping wings in Sherwood-Relics may be too strong when compared with the observations. Indeed, when comparing the best fit ionized bubble sizes from \citet{umeda2023} with the median bubble sizes around the hundred most massive haloes in Sherwood-Relics, we find that at $z=7.140 \, (7.452)$ the bubble sizes are a factor 13 (11) smaller than the measurements from \citet{umeda2023}. The large bubble sizes measured in that work are driven by the significant detections of transmitted flux in pixels blueward of the Ly$\alpha$ source frame wavelength. Fitting for this with the \citet{miraldaescude1998} model for the IGM damping wing allows for this detected flux when the ionized bubbles around the host galaxies are large. 

However, based on the results of Section \ref{section:compare_model}, we do not expect to observe any transmitted flux blueward of the Ly$\alpha$ source frame wavelength, independent of the size of the ionized bubble, as a result of the residual neutral hydrogen in the ionized IGM. Furthermore, if there were indeed large, 100 cMpc-scale regions of the IGM that were transparent to Ly$\alpha$ scattering already around $z \sim 7$ galaxies, one would expect to see these transmissive regions in the $z>6.5$ Ly$\alpha$ forest of the highest redshift quasars, but this is not the case \citep{jin2023}. An exception would be the highly ionized proximity zones observed around high-redshift quasars, but even around the brightest $z\sim 6$ quasars, the sizes of these proximity zones are smaller than the sizes of ionized bubbles measured by \citet{umeda2023} at $z \sim 7$ \citep{eilers2017}. The amplitude of the transmitted Ly$\alpha$ forest flux proposed by \citet{umeda2023} is also unexpected. Using their best-fit parameters in the \citet{miraldaescude1998} model at $z=7.140$, we can measure the mean flux in the galaxy's Ly$\alpha$ forest in two 50 $h^{-1}$ cMpc chunks that contain sections of the ionized bubble. We measure a mean flux $\langle F \rangle$ = (0.97, 0.69) in the two chunks, as we move blueward from the galaxy. This can also be expressed as an effective optical depth $\tau_{\rm eff}$ = (0.03, 0.37). Values as high as this are not observed in the Ly$\alpha$ forest of quasars until below $z =3$ \citep{becker2013}. 

It is therefore of interest to investigate other scenarios that could produce transmitted flux blueward of the galaxy's Ly$\alpha$ emission. One possibility is that the flux and wavelength calibration of JWST PRISM spectra still present challenges. Residual errors from these may result in additional spurious flux blueward of Ly$\alpha$.
The choice of the intrinsic galaxy spectrum may also play a role, and this possibility is discussed in Section \ref{sec:spectrum} below.

\subsection{Influence of the intrinsic spectrum on inferred bubble sizes}
\label{sec:spectrum}

An alternative explanation for these observations may arise from the chosen intrinsic galaxy template spectrum. If the assumed intrinsic spectrum used is not representative of the true intrinsic spectrum, this may lead to a change in the reionization parameters recovered from the observations. In particular, an underestimate of the strength the Ly$\alpha$ emission line will effect not just the pixel that line falls in, but also the neighbouring pixels. This is a result of the low resolution of the NIRSpec PRISM mode, which has $R = \frac{\Delta\lambda}{\lambda} \sim 30-40$ in the redshift range $z = 7-10$. We investigate the effect of underestimating the intrinsic Ly$\alpha$ emission of the galaxy in the right column of Figure \ref{fig:umeda23_stacks}. We repeat the process of convolving the Sherwood-Relics IGM damping wings with a template galaxy spectrum. We again assume the same stacks of lower redshift spectra from \citet{cullen2019}, but we now adjust the strength of the Ly$\alpha$ line such that it is a factor of three stronger. 

This factor of three was chosen arbitrarily as a value to demonstrate the effect of changing the intrinsic spectrum. The Ly$\alpha$ equivalent width distribution in the sample used to make the low-redshift stacks has a large scatter, with not all galaxies showing Ly$\alpha$ emission and some showing absorption instead. In \citet{cullen2020}, which analysed an expanded sample compared to \citet{cullen2019}, the median rest-frame equivalent width of the sample was $W_{\rm Ly\alpha} = -4$ \AA. The strongest Ly$\alpha$ emitting galaxies had $W_{\rm Ly\alpha} = 110$ \AA. \citet{shibuya2018} presented observations of $z\sim6.6$ galaxies with equivalent widths $W_{\rm Ly\alpha} > 200$ \AA. These observations will also contain the effects of attenuation by the IGM, so the intrinsic Ly$\alpha$ emission of these galaxies will be stronger. Multiplying the Ly$\alpha$ emission in these stacked spectra by a factor of three to produce intrinsic Ly$\alpha$ equivalent widths of at most $W_{\rm Ly\alpha} \sim 330$ \AA \, in individual galaxies, and lower values in the stacked spectra, is therefore not totally unreasonable. Indeed, a Ly$\alpha$ emitting galaxy with $W_{\rm Ly\alpha} \approx 400$ \AA \, at $z = 7.278$ has been discovered with JWST \citep{saxena2023}. We note that we only increase the amplitude of the line, and do not change its shape, but this should not make a difference given the wide pixels used in the stacked spectra. 

The resulting template is shown by the thin purple lines in Figure \ref{fig:umeda23_stacks} before convolution and binning, and by the thick purple curves afterwards. Comparing the thick orange and thick purple curves in the figures on the left and right columns, we find that, as expected, increasing the intrinsic Ly$\alpha$ emission enhances the intrinsic flux not only in the pixel that corresponds to the source frame Ly$\alpha$ wavelength, but also in adjacent pixels, due to the broad instrument profile of the NIRSpec PRISM mode \citep[see also][]{jones2023}. We further find that we now see an excess of flux in the pixel containing the Ly$\alpha$ emission line in our template intrinsic spectrum compared to the observations at $z=7.140$ and 7.452. This is in contrast to the template used in \citet{umeda2023}, where the expected and observed flux at Ly$\alpha$ were nearly identical, due to the large ionized bubbles that were assumed. If this latter option were the case, it would be a different scenario from what is observed in the spectra of $z=7-7.5$ quasars, where a significant fraction of the Ly$\alpha$ emission line is expected to be absorbed \citep[e.g.,][]{davies2018,wang2020,greig2022}.

We find that when we convolve the Sherwood-Relics IGM damping wings with this new template for the intrinsic galaxy spectrum, we now find significant transmitted flux blueward of the Ly$\alpha$ source frame wavelength, mimicking the effect of invoking the large ionized bubbles in the \citet{miraldaescude1998} model. This effect is largest in the lowest redshift bins, as the IGM damping wing in the simulations is strong enough at $z \gtrsim 8$ to absorb the bulk of the Ly$\alpha$ emission. Of course, we have here arbitrarily changed the strength of the intrinsic Ly$\alpha$ emission by hand to demonstrate the expected effect. In practice, it should be possible to obtain an estimate of the expected intrinsic Ly$\alpha$ emission from the Balmer emission lines of the galaxy \citep[e.g.,][]{hayes2015}.

We further note that we found near-indistinguishable results when comparing the \citet{umeda2023} observations to the Sherwood-Relics late and early models (plotted in Figure \ref{fig:HI_history}). Although there are differences in the damping wings we compute from these two simulations, as shown in Figure \ref{fig:compare_model_history_boxsize}, the broad instrument profile of the NIRSpec PRISM mode and large pixel size of the stacked spectra makes the two models difficult to differentiate when comparing to the JWST data, and the change in reionization history was much less significant than the change in the intrinsic spectrum. This suggests that it may be difficult to constrain reionization with the currently published spectra, and that larger samples of galaxies will be required.

\subsection{Proximate high column density absorbers}
\label{sec:compare_DLAs}

\begin{figure}
\includegraphics[width=\columnwidth]{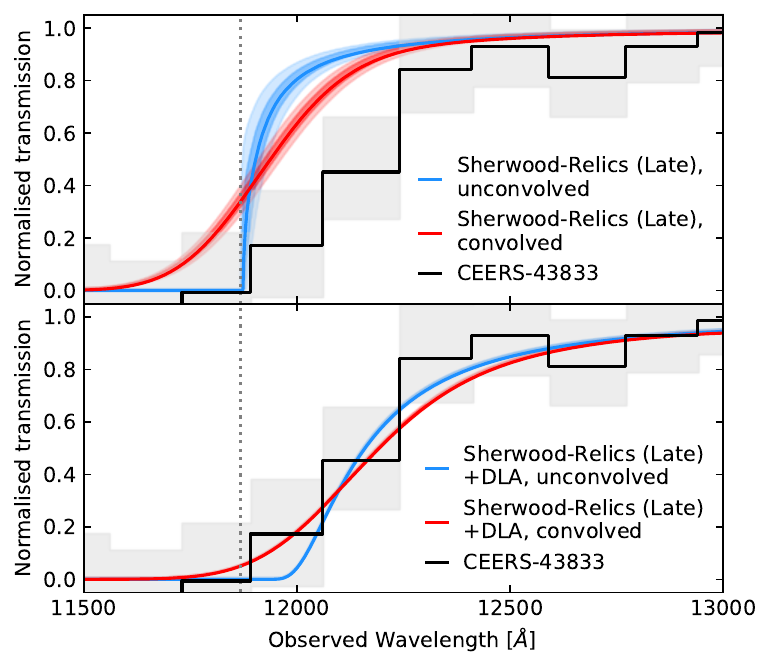}
    \caption{Comparison of the normalised spectrum of CEERS-43833 (a galaxy at redshift $z=8.7622$) taken from \citet{heintz2023} and the IGM damping wings from the Sherwood-Relics late reionization simulation. The black line in both panels shows the spectrum of CEERS-43833 and the grey shaded regions show the corresponding error. \textit{Top}: The blue line and shaded regions are the median and scatter in IGM transmission curves calculated from the simulation. The red line and shaded regions are the same curves, but after convolution with the NIRSpec PRISM instrument profile. \textit{Bottom}: As above, but now the simulated damping wings are computed for both the contribution from the IGM and for a proximate DLA with column density $N_{\rm HI} = 10^{22.1}$ cm$^{-2}$, as estimated by \citet{heintz2023}.}
\label{fig:ceers_spec}
\end{figure}

We next analyse our simulated IGM damping wings in the context of the detection of damped Lyman-$\alpha$ absorbers in three galaxies at $z=8-11$ \citep{heintz2023}. The spectra of these three galaxies were compared with IGM damping wings calculated using the \citet{miraldaescude1998} with volume-weighted average neutral fraction $\langle x_{\rm HI} \rangle_{\rm v} = 0.1, 0.5$ and $0.1$ and assuming an ionized bubble radius $R_{\rm b} = 0$ (i.e., setting $z_{\rm b}$ equal to $z_{\rm s}$ in Equation \ref{eqn:DW}). Under these assumptions, it was shown that IGM damping wing profiles were not strong enough to explain the observed damping wings. Further comparison with simulated IGM damping wings from \citet{laursen2019} also indicated that the IGM alone could not explain the observations. However, as noted in that work, the neutral fraction in that simulation is somewhat low, with $\langle x_{\rm HI} \rangle_{\rm v} = 0.13$ at $z=8.8$. \citet{heintz2023} concluded that including strong proximate DLAs with \ion{H}{i} column densities in the range $N_{\rm HI} = 10^{22.1}-10^{22.4} \, \rm{cm}^{-2}$ provided a much better fit to the observations. 

We test this theory in the context of the IGM transmission curves generated from the Sherwood-Relics late reionization model, which as described in Section \ref{section:model_diff} contain several physical effects that are not present in the \citet{miraldaescude1998} model and which has a reionization history in good agreement with estimates of the evolution of the IGM neutral fraction (Figure \ref{fig:HI_history}). We note that we do not expect to produce DLAs in our simulations, due to our simplified star formation model and lack of galactic feedback. We further note that, as described in Section \ref{section:describe_spectra}, we are also not including the contribution of any gas inside the virial radius of the galaxy when we compute our IGM damping wings. We focus our comparison on CEERS-43833, a galaxy at redshift $z=8.7622$ \citep{arrabalharo2023,finkelstein2023}, as \citet{heintz2023} show a version of this spectrum where the transmission has been normalised by the intrinsic transmission, and so can be easily compared with our simulated damping wings. 

We compute our absorption spectra starting from haloes extracted from a simulation snapshot at $z=9$, but rescale the gas density and velocity assuming the host halo is at the redshift of CEERS-43833. The simulated IGM transmission curves we calculate are shown in blue in the top panel of Figure \ref{fig:ceers_spec} and can be compared with the normalised spectrum of CEERS-43833 shown in black. As in Section \ref{sec:compare_umeda23}, we further convolve the spectra with a Gaussian with a width corresponding to the resolution at the Ly$\alpha$ wavelength at the redshift of CEERS-43833. The resulting transmission curves are shown in red in Figure \ref{fig:ceers_spec}. We find that even after convolution, the absorption from the IGM alone is not enough to reproduce the damping wing observed in CEERS-43833. This is similar to the simulated IGM damping wings of \citet{laursen2019}, despite the later reionization model we analyse here. This supports the claim of \citet{heintz2023} that there are proximate DLAs associated with these high-redshift star-forming galaxies. Indeed, when we include the contribution of a DLA with column density $N_{\rm HI} = 10^{22.1} \, \rm{cm}^{-2}$ at the redshift of the host galaxy, we find a much better fit to the observed spectrum, as shown in the bottom panel of Figure \ref{fig:ceers_spec}.

\subsection{Implications}

It is clear from these JWST observations that IGM damping wings seen in high-redshift galaxies open up many exciting possibilities for studying the first half of the reionization epoch. However, this observable comes with its own set of challenges. Galaxy spectra are expected to be somewhat simpler to model than quasar spectra as, e.g., the quasar spectra can be complicated by strong \ion{N}{v} emission close to the Ly$\alpha$ emission \citep{davies2018b, greig2022}. However, the strength of the intrinsic Ly$\alpha$ emission will still be important to capture, especially in analyses of low resolution spectra. Moving forward, it may be useful to marginalise over grids of galaxy template spectra convolved with a separate grid of IGM damping wing models, in a spirit similar to what has been carried out in analyses of quasar damping wings. However, as pointed out by \citet{umeda2023}, it will be necessary to understand how to incorporate absorption by the circumgalactic medium of the host galaxy in such a setup.

The incidence rate of proximate DLAs in these galaxies will also add a source of uncertainty. Indeed, two of the three galaxies shown to host DLAs in \citet{heintz2023} (MACS0647-JD at $z=10.170$ and CEERS-16943/Maisie's Galaxy at $z=11.409$) are also present in the sample analysed by \citet{umeda2023}. In the future, it will be important to either allow for the contribution of local, high-density neutral hydrogen to the observed damping wings when fitting for the IGM component, as has been done for GRBs \citep{totani2006}. Alternatively, it may be possible to remove the contaminated objects from the sample, by searching for metal lines redward of Ly$\alpha$ that are associated with the DLA. However, depending on the metallicity of these objects, this may have to wait for the upcoming spectrographs on 30-metre class telescopes, such as ANDES or HARMONI. It may also be possible to remove the DLA hosts statistically based on the properties of the galaxy, once we have accumulated a large enough sample of spectra from these $z > 7$ galaxies.
 
At the same time, it will be important to continue to make advances on the theoretical side to properly interpret these results. For example, it is likely that these simulations would struggle to reproduce observations of the (rare) double-peaked Ly$\alpha$ emitting galaxies at $z \sim 6.5$ \citep{hu2016,songaila2018,meyer2021}, which may call for more efficient production and escape of ionizing photons than we have assumed in the source model for our simulations or the contribution of AGN in locally enhancing the ionization state of the IGM \citep{bosman2020}. It will further be important to increase the dynamic range of the simulations, such that larger volumes can be modelled while still resolving the small-scale structure of the IGM, to bridge the gap between our models and the bubble sizes predicted in large semi-numerical simulations of reionization \citep{lu2023} to fully understand the implications for Ly$\alpha$ transmission.

\section{Conclusions}
\label{section:conclusions}

We have presented here an analysis of mock IGM damping wings generated from the Sherwood-Relics simulation suite. We have compared our simulated damping wings to the analytic IGM damping wing model of \citet{miraldaescude1998}. We found excellent agreement between the simulated and analytic damping wings redward of Ly$\alpha$, but found that the agreement was poor on the blue side of Ly$\alpha$, with the analytic model predicting much more IGM transmission than expected from the simulated damping wings. We showed that this was a result of the residual neutral hydrogen inside in the ionized bubbles in our simulations of inhomogeneous reionization, which is enough to saturate the IGM absorption.

We further compared the simulated damping wings against recent observations of damping wings in high-redshift galaxies seen with JWST. We found that the simulated damping wings were unable to reproduce the significant detections of transmitted flux blueward of the galaxy Ly$\alpha$ emission, which have led to claims of 100 cMpc-sized ionized bubbles at $z\sim7-7.5$. We suggest that as the residual neutral gas in the IGM at $z\sim 7$ is expected to saturate the IGM absorption, the flux that has been observed blueward of Ly$\alpha$ may instead be a result of stronger intrinsic Ly$\alpha$ emission, which is spread into nearby pixels by the instrument profile of the NIRSpec PRISM mode. We also investigated claims of observed proximate DLAs in high-redshift galaxies, and confirmed that the strong damping wings observed in these galaxies cannot be reproduced by the IGM damping wings generated from our simulations.

Of course, we are only witnessing the first attempts at constraining the first half of reionization with galaxy damping wings. The first results are already incredibly impressive, and show huge promise for revealing the timing of reionization, even before the presence of this neutral gas is directly detected through its 21cm transition in upcoming experiments. However, this work has demonstrated that the exquisite quality of the new JWST data demands comparison with models that include all of the relevant physics. Future progress in this area will come from advances on both the observational and theoretical sides in understanding the interactions between galaxies and the intergalactic medium in the reionization epoch.

\section*{Acknowledgements}

The simulations used in this work were performed using the Joliot Curie supercomputer at the Tr{\'e}s Grand Centre de Calcul (TGCC) and the Cambridge Service for Data Driven Discovery (CSD3), part of which is operated by the University of Cambridge Research Computing on behalf of the STFC DiRAC HPC Facility (www.dirac.ac.uk).  We acknowledge the Partnership for Advanced Computing in Europe (PRACE) for awarding us time on Joliot Curie in the 16th call. The DiRAC component of CSD3 was funded by BEIS capital funding via STFC capital grants ST/P002307/1 and ST/R002452/1 and STFC operations grant ST/R00689X/1.  This work also used the DiRAC@Durham facility managed by the Institute for Computational Cosmology on behalf of the STFC DiRAC HPC Facility. The equipment was funded by BEIS capital funding via STFC capital grants ST/P002293/1 and ST/R002371/1, Durham University and STFC operations grant ST/R000832/1. DiRAC is part of the National e-Infrastructure.  JSB is supported by STFC consolidated grant ST/T000171/1. FC acknowledges support from a UKRI Frontier Research Guarantee Grant (PI Cullen; grant reference EP/X021025/1). MGH is supported by STFC consolidated grant ST/S000623/1. GK is partly supported by the Department of Atomic Energy (Government of India) research project with Project Identification Number RTI~4002, and by the Max Planck Society through a Max Planck Partner Group. Support by ERC Advanced Grant 320596 ‘The Emergence of Structure During the Epoch of Reionization’ is gratefully acknowledged. We thank Kasper Heintz for answering a question about the spectrum of CEERS-43833. We thank Volker Springel for making \textsc{p-gadget-3} available. We also thank Dominique Aubert for sharing the \textsc{aton} code. For the purpose of open access, the author has applied a Creative Commons Attribution (CC BY) licence to any Author Accepted Manuscript version arising from this submission.

\section*{Data Availability}

All data and analysis code used in this work are available from the first author on reasonable request.



\bibliographystyle{mnras}
\bibliography{ref} 


\bsp	
\label{lastpage}
\end{document}